\documentclass[conference,compsocconf]{IEEEtran}

\ifCLASSINFOpdf
\else
\fi
\usepackage{graphicx,color,colortbl,psfrag,subfigure}
\usepackage[usenames,dvipsnames]{xcolor}
\usepackage{bbding}
\usepackage{subfigure}
\usepackage{tabularx}
\usepackage{multirow}
\usepackage{array}
\usepackage{upgreek}
\usepackage{bigints}
\usepackage{url}

\usepackage{epsfig}
\usepackage{mdwlist}
\usepackage{array}
\usepackage{bm}

\usepackage{amssymb}
\usepackage{amsmath}
\usepackage{amsfonts}

\usepackage[thmmarks,thref,hyperref,amsmath]{ntheorem}

\usepackage{algpseudocode}
\usepackage{algorithm}

\usepackage[first=0,last=9]{lcg}

\definecolor{Gray}{gray}{0.7}

\begin{document}
%
\title{Risk-Aware Dynamic Reserve Prices of Programmatic Guarantee in\\ Display Advertising}


\author{\IEEEauthorblockN{Bowei Chen}
\IEEEauthorblockA{
School of Computer Science\\ 
University of Lincoln\\
Lincoln, United Kingdom\\
bchen@lincoln.ac.uk}
}


%

\maketitle

\begin{abstract}

Display advertising is an important online advertising type where banner advertisements (shortly ad) on websites are usually measured by how many times they are viewed by online users. There are two major channels to sell ad views. They can be auctioned off in real time or be directly sold through guaranteed contracts in advance. The former is also known as real-time bidding (RTB), in which media buyers come to a common marketplace to compete for a single ad view and this inventory will be allocated to a buyer in milliseconds by an auction model. Unlike RTB, buying and selling guaranteed contracts are not usually programmatic but through private negotiations as advertisers would like to customise their requests and purchase ad views in bulk. In this paper, we propose a simple model that facilitates the automation of direct sales. In our model, a media seller puts future ad views on sale and receives buy requests sequentially over time until the future delivery period. The seller maintains a hidden yet dynamically changing reserve price in order to decide whether to accept a buy request or not. The future supply and demand are assumed to be well estimated and static, and the model's revenue management is using inventory control theory where each computed reverse price is based on the updated supply and demand, and the unsold future ad views will be auctioned off in RTB to the meet the unfulfilled demand. The model has several desirable properties. First, it is not limited to the demand arrival assumption. Second, it will not affect the current equilibrium between RTB and direct sales as there are no posted guaranteed prices. Third, the model uses the expected revenue from RTB as a lower bound for inventory control and we show that a publisher can receive expected total revenue greater than or equal to those from only RTB if she uses the computed dynamic reserves prices for direct sales. 

\end{abstract}

\begin{IEEEkeywords}
Display advertising; programmatic guarantee; dynamic inventory control; risk-aware modelling; reserve prices
\end{IEEEkeywords}

%
\IEEEpeerreviewmaketitle


\section{Introduction}
\label{sec:introduction}

Display advertising -- a type of online advertising that mainly comes with banner ads to deliver marketing messages to site visitors -- has emerged as a new global industry as billions of dollars are spent every year for ad views. Each ad view is also called an \emph{impression}. According to eMarketer\footnote{\url{http://www.emarketer.com}}, display advertising revenues mainly come from two channels: RTB and direct sales. The former, as the name implies, is a real-time, impression-level, auction-based sales system, in which media buyers come to a common marketplace like ad exchange to compete with each other for a single impression from their targeted users. RTB was initially proposed in 2007~\cite{Google_2011}, which has brought automation, integration, and liquidity into selling impressions non-guaranteedly, changing the landscape of the market. Direct sales have a longer history, which can be traced back to 1994, when HotWired (today Wired News, part of Lycos) singed fourteen banner ads with AT\&T, Club Med and Coor’z Zima, being considered as the start of display advertising~\cite{Bruner_2005}. Guaranteed contracts can customise media buyers' requests and provide a way to lock in advertising opportunities in advance. However, they are still mainly agreed through private negotiations, which is slow and less efficient in front of large inventory volumes and rapid market changes. Both media buyers and sellers are looking to programmatic technology or automatic system to buy and sell impressions in advance, bypassing traditional direct sales~\cite{OpenX_2013}.

Programmatic guarantee (PG) has therefore become a popular topic recently. It is often synonymous with programmatic reserved, programmatic upfront and forward~\cite{OpenX_2013,Dunaway_2012}. Essentially, PG is a sales system that helps media sellers such as publishers and supply-side platforms (SSPs) to automatically sells future impressions through guaranteed contracts to media buyers such as advertisers and demand-side platforms (DSPs). Below are several notable examples from both buy-side and sell-side markets:
\begin{itemize}
\item Google DoubleClick's Programmatic Guaranteed;
\item AOL's Programmatic Upfront;
\item Rubicon Project's Reserved Premium Media Buys;
\item BuySellAds's Direct Sales;
\item ShinyAds's Programmatic Direct Advertising Platform.
\end{itemize}

\begin{figure}[t]
\centering
\includegraphics[width=1\linewidth]{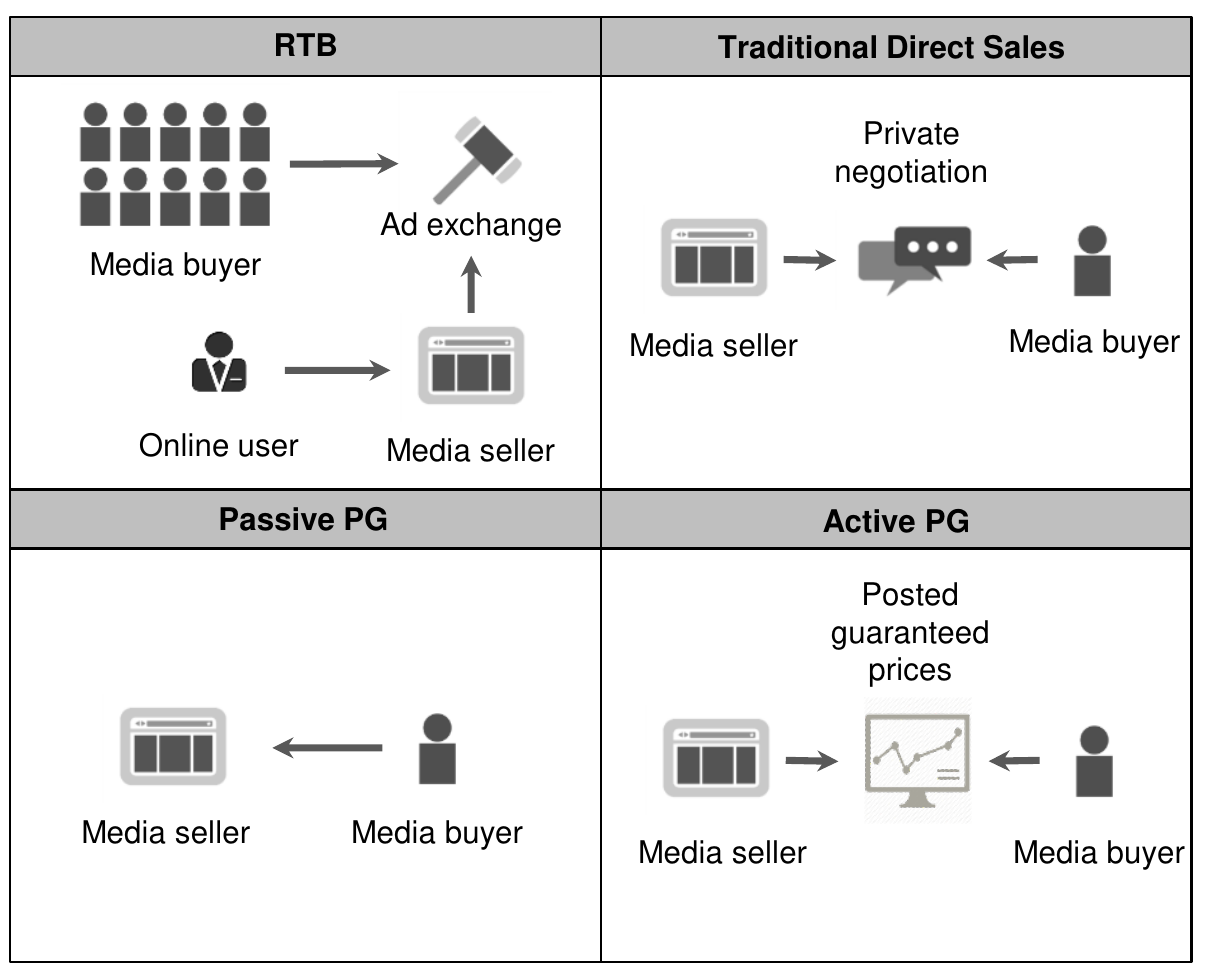}
\caption{Schematic views of transactions (or business models) in RTB, traditional direct sales, passive PG and active PG.}
\label{fig:schematic_view}
\end{figure}

Many PG products or platforms have been developed since 2013 and there is no widely recognised models for them. This is different to RTB, where the Second-Price (SP) auction and the Vickrey-Clarke-Groves (VCG) auction have been implemented on most of platforms~\cite{Varian_2009}. In addition, since in RTB each impression is auctioned off, the expected revenues from SP and VCG are equivalent~\cite{Parkes_2007}; therefore, we usually consider RTB based on SP auctions. Below we summarise some functionalities and common facts of PG. First, there is no competition in PG. For each transaction, there are only one seller and one buyer at a time. Second, there is no need to model the arrival of supply in PG as the inventory is the impressions that will be created in the future period but not now. However, the prediction of supply in the future period is important. Third, the arrival of demand needs to be considered, no matter whether it is considered as a queue or shifting from future prediction. Fourth, there is no negotiation process in PG. In its design, either a buyer submits her request or a seller posts the guaranteed inventory price publicly. In this paper, we call the former the \emph{passive PG} and the later the \emph{active PG}. A passive PG usually comes with hidden reserve prices to automatically accept the buyer's request. Many current PGs adopt this design. An active PG is similar to an airline booking system~\cite{Talluri_2004}, in which media buyers can monitor the evolution of the guaranteed prices and then adjust their advertising strategies between RTB and PG. This will also affect the current balance of the two markets but will achieve a new equilibrium by using the posted prices to affect advertisers' demand. For the reader's convenience, the key components of the mentioned display advertising sales systems are summarised in Fig.~\ref{fig:schematic_view}. 

In this paper, we discuss a simple dynamic model for a passive PG where a publisher allows advertisers to submit guaranteed buy requests to purchase their targeted future impressions in advance. The model calculates a hidden reserve price to decide whether to accept a buy request or not. To simplify the discussion without loss of generality, we consider each buy request contains a single impression and requests arrive one by one over time (as a queue). The reserve price calculation is based on the updated dual force of supply and demand, and the unsold impressions will be auctioned off in RTB to the meet the unfulfilled demand in the delivery period. The model can be easily applied to the case of bulk sales. Inventory control theory has been employed for revenue management and we use the expected revenue from future RTB as a lower bound. The model is not optimal while it still has several desirable properties. Firstly, it is not limited to the demand arrival assumption because there are not posted prices to affect demand, and the model also doesn't need to find a global optimality for revenue maximisation. Second, it will not affect the current equilibrium between RTB and direct sales because the reserve prices are not disclosed. Third, we show that a seller can receive an increased expected total revenue compared to RTB if she uses the computed dynamic reserves prices for her direct sales. 

The rest of the paper is organised as follows. Section~\ref{sec:related_work} reviews the related literature. In Section~\ref{sec:the_model}, we formulate the problem, discuss assumptions and provide a solution. Section~\ref{sec:experiments} presents the results of our experimental evaluation and Section~\ref{sec:conclusion} concludes the paper.

\section{Related Work}
\label{sec:related_work}

Several recent developments of guaranteed delivery systems in both display advertising and sponsored search are reviewed in this section. 

The allocation of impressions between guaranteed and non-guaranteed channels was investigated through various approaches. Feldman et al.~\cite{Feldman_2009} studied an algorithm that can allocate and match ads for display advertising, in which the publisher\rq{}s objective is not only to fulfil the guaranteed contracts but also to deliver the well-targeted impressions to advertisers. The algorithm allows for free disposal so that advertisers are indifferent to, or prefer being assigned more than a certain number of impressions without changing the contract terms. Ghosh et al.~\cite{Ghosh_2009} considered the publisher as a bidder to bid for guaranteed contracts so that impressions would be possible allocated to auctions only if the winning bids are high enough. Balseiro et al.~\cite{Balseiro_2011} used stochastic control theory to  model the decision-making in the same scenario. Roels and Fridgeirsdottir~\cite{Roels_2009} further proposed several control heuristics in revenue maximisation. Given static supply and demand, optimal posted guaranteed prices were recently discussed by Chen et al.~\cite{Chen_2014_2}.

Other related contributions include: a lightweight allocation framework for guaranteed impressions that simplifies the computations in optimisation and to let real servers to allocate ads efficiently and with little overhead~\cite{Bharadwaj_2012}; two contract pricing algorithms to calculate the price of selling guaranteed impressions in bulk~\cite{Bharadwaj_2010}; guaranteed delivery mechanisms with cancellations (for media seller)~\cite{Babaioff_2009,Constantin_2009} 

%
 
Ad options are a special type of guaranteed delivery systems worthing to be mentioned~\cite{Moon_2010,Wang_2012_1,Chen_2015_1,Chen_2015_2}, in which an advertiser is guaranteed a priority buying right (but not obligation) of her targeted future inventories. She usually pays a small amount upfront and then can decide to pay a fixed price to obtain the inventories or not in the delivery period or contract expiration date. She can join keyword auctions or RTB in the future if she thinks they are less expensive. The only cost would be the prepaid option price. Compared to guaranteed contracts, ad options provide advertisers with great flexibility.


\begin{figure}[t]
\centering
\includegraphics[width=1\linewidth]{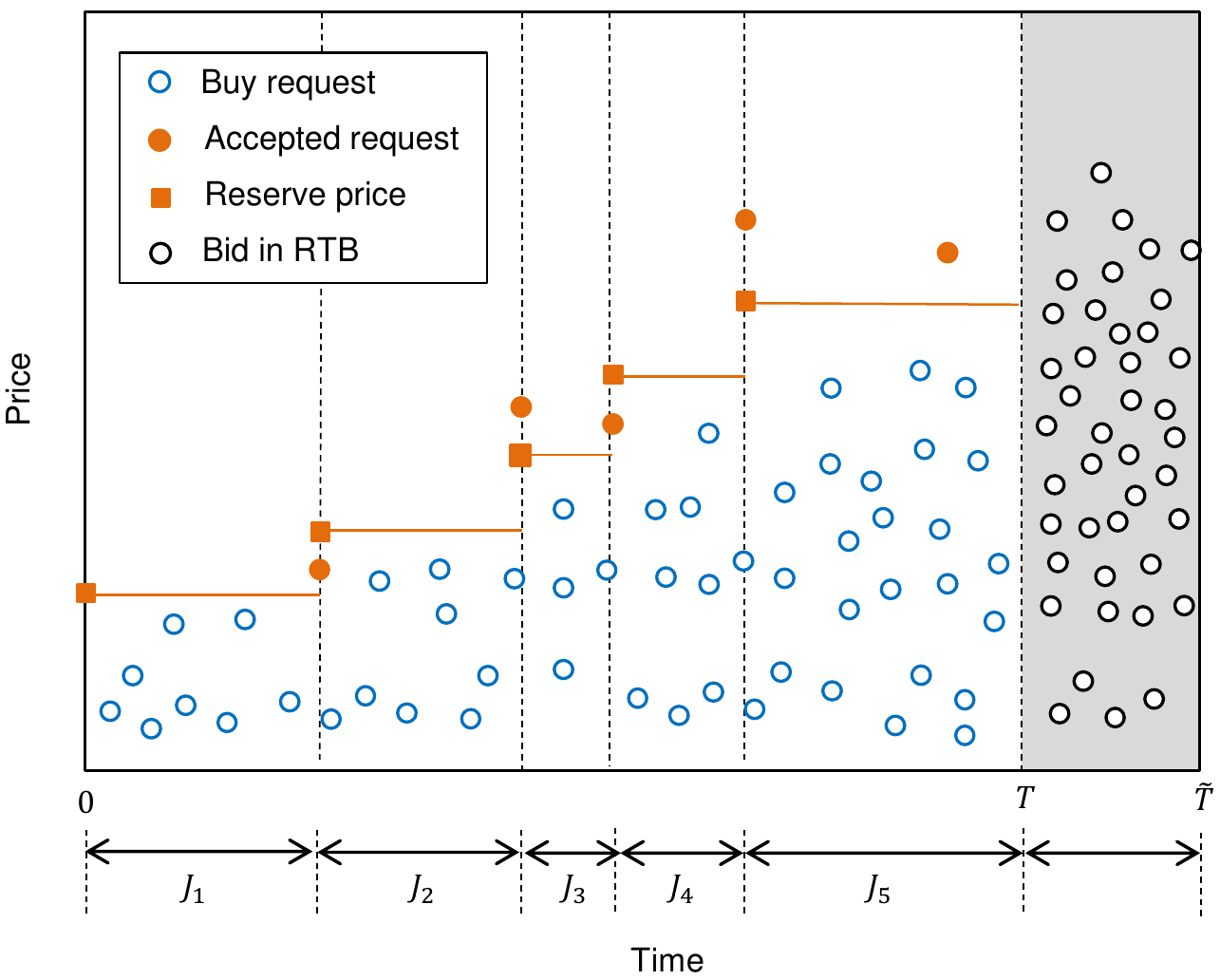}
\hspace{-5pt}
\caption{A schematic illustration of using risk-aware dynamic reserve prices to buy and sell guaranteed impressions in display advertising.}
\label{fig:dynamic_framework}
\end{figure}

\section{The Model}
\label{sec:the_model}

In display advertising, a publisher may have several webpages. On a webpage, there exists one or more slots (or placements) to display banner ads. When an online user visits a publisher's webpage, an ad slot can generate a chance of ad view (i.e, the impression), which is usually auctioned off in RTB, where advertisers compete in a SP auction and the winner will be able to have her ad displayed to the user. This impression can also be sold in advance through a guaranteed contract and our proposed model brings automation into selling guaranteed impressions. To simplify the discussion without loss of generality, we consider a single publisher, single webpage, single slot, and single impression at a time in the modelling. It should be noted that guaranteed impressions are usually sold in bulk in real businesses and our model can be easily applied to the bulk sales case. The model can also be used by a SSP for selling premium impressions and/or from a specific user group.

Fig.~\ref{fig:dynamic_framework} presents a simple schematic illustration of our model. We assume the targeted impressions will be created in the period $[T,\widetilde{T}]$, and they can also be sold in advance through guaranteed contracts in the period $[0,T]$. Advertisers can submit buy requests one by one (as a queue) and each request is for a single impression. The publisher will decide whether to accept or reject a buy request based on a reserve price, which is determined by the updated information of dual force of supply of and demand for impressions in the period $[T, \widetilde{T}]$. Finally, the remaining impressions and the unfulfilled demand will go to RTB in the period $[T,\widetilde{T}]$. Here we use $J_1,J_2,\cdots$ to denote discrete time periods for guaranteed buy requests and each period contains only one accepted buy request. 

\subsection{Dynamic Decision Making}
\label{sec:decision_making}

Let $V(t,s)$ be the publisher's value function at time $t$, representing the expected total value of $s$ remaining impressions which will be created and delivered in the future period $[T, \widetilde{T}]$. Suppose if an advertiser submits a buy request to the publisher and proposes a guaranteed price $G(t)$ for an impression for targeted users, the publisher's decision making at time $t$ can be simply expressed as
\[
\max_{x(t) \in \{0, 1\}} 
\Big\{ R(t) x(t) + V\big(t+\delta t, s-x(t)\big) \Big\},
\]
where $x(t)$ is the decision variable with binary outcomes, $R(t)$ is the expected revenue that can be obtained. Therefore, given a buy request, the publisher's decision making will be based on the maximisation of the sum of the current expected revenue and the expected revenue of future. As the publisher may fail to deliver the guaranteed impression in the future, her expected penalty needs to be considered. Let $\omega$ be the probability that the publisher fails to deliver the guaranteed impression in the delivery period and let $\gamma$ be the size of penalty -- if the publisher fails to deliver the guaranteed impression that is sold at $G(t)$, she needs to pay $\gamma G(t)$ penalty to the advertiser. 

Given time $t$ and $s$ remaining impressions, suppose that the advertiser proposes a guaranteed price that makes the publisher's two decisions \emph{indifferent} -- choosing either will give her equivalent benefits. We consider this price the lower bound of \emph{reserve price} for the guaranteed impression and denote it by $r(t,s)$ as the price will be affected by both time and remaining impressions. Mathematically, $r(t,s)$ can be expressed as
\begin{equation}
\label{eq:reserve_price}
r(t,s) = 
\frac{1}{1 - \gamma \omega} 
\Big(V(t + \delta t, s) - V(t + \delta t, s-1)\Big).
\end{equation}
Then the decision variable $x(t) = \mathbb{I}_{\{G(t) \geq r(t, s)\}}$, where $\mathbb{I}_{\{\cdot\}}$ is the indicator function. Applying the Bellman's Principle of Optimality~\cite{Talluri_2004} then gives 
\begin{align*}
  & \ V(t, s) \nonumber \\
= & \ \mathbb{E} \Big[\max_{x(t)} \Big\{ r(t, s) (1-\gamma \omega ) x(t) + V(t+\delta t, s - x(t)) \Big\} \Big] \nonumber \\
= & \ \mathbb{P}\big[G (t) \geq r(t, s)\big] \Big( r(t, s) (1-\gamma \omega ) + V(t+ \delta t, s - 1)\Big) \nonumber \\
   & \  + \Big( 1 - \mathbb{P}\big[G(t) \geq r(t, s)\big] \Big) V(t + \delta t, s). 
\end{align*}


In the same way, $V(t + \delta t, s)$ can be obtained. By Eq.~(\ref{eq:reserve_price}), $r(t+\delta t, s)$ is defined, then $V(t +\delta t, s) = V(t + 2 \delta t, s)$. Substituting the publisher's value functions into Eq.~(\ref{eq:reserve_price}) gives $r(t +\delta t, s) = r(t + 2 \delta t, s)$. Same checking can be applied to multiple steps, and we then find that the publisher's value function and the reserve price are both time independent. For any $k \in [t, T]$, $V(t, s) = V(k, s), r(t, s) = r(k, s)$. Hence, if the publisher keeps selling impressions at the lower bounds of reserve prices, her expected marginal revenues are always equal to her expected marginal costs, and her expected total value will be keeping at the same level up to the terminal time $T$.



\subsection{Terminal Value}
\label{sec:terminal_value}

Let $S$ and $Q$ be the expected total supply of and demand for impressions that will be created in the period $[T, \widetilde{T}]$, respectively. Consider if $S-s$ impressions have been sold in advance through guaranteed contracts in the period $[0, T]$ and there are $s$ remaining impressions which will be auctioned off in RTB in the period $[T, \widetilde{T}]$. Recall that the sold impressions have also fulfilled $S-s$ demand and unfulfilled $Q - (S - s)$ demand will join RTB. The publisher's value function at time $T$, also called \emph{terminal value}, can be obtained by $V( T, s ) = s \phi (\xi)$, where $\xi$ is the per-impression demand (i.e., the number of advertisers) and $\phi( \cdot )$ is the function which computes the estimated per-impression payment price in RTB for the given demand level. Since there are $s$ remaining impressions and $Q - (S - s)$ remaining demand, then $\xi = (Q-S)/s + 1$. 

As advertisers usually bid for impressions separately in RTB~\cite{Google_2011}, it can be considered as a single-item auction where GSP and VCG auction models have equivalent expected revenues. Therefore, $\phi(\cdot)$ can be obtained as follows 
\begin{align*}
\phi(\xi) = & \ \int_{x \in \Omega} \hspace{-5pt} x \xi (\xi - 1) g(x) \big(1 - \mathbb{F}(x) \big) \big(\mathbb{F}(x)\big)^{\xi-2} dx, 
\end{align*}
where $x$ is an advertiser's bid in RTB, $g(\cdot)$ is the density function, $\mathbb{F}(\cdot)$ is the cumulative distribution function, so $\xi (\xi - 1) g(x) (1 - \mathbb{F}(x) ) (\mathbb{F}(x))^{\xi-2}$ represents the probability that if an advertiser who bids at $x$ is the second highest bidder, then one of $\xi - 1$ other advertisers must bid at least as much as she does and all of $\xi - 2$ other advertisers have to bid no more than she does.

Bid distribution can be specified by either probabilistic or empirical method. In probabilistic way, uniform distribution and log-normal distribution have been widely discussed for online advertising auctions~\cite{Ostrovsky_2011,Milgrom_1982,Menezes_2008}. If bid $X \sim \mathbf{U}[0,v]$, where $v$ is the advertiser's expected value on an impression, then $\phi(\xi) = v (\xi - 1)/(\xi + 1)$. If bid $X \sim \mathbf{LN}(\mu, \sigma^2)$, where $\mu$ and $\sigma$ are mean and standard deviation, then $\phi(\xi)$ can be obtained via numerical integration. Probabilistic methods offer many statistical properties while they are not valid empirically in many situations~\cite{Chen_2014_2,Yuan_2014,Chen_2015_2}. In this paper, we learn $\phi(\cdot)$ from data empirically by using the robust locally weighted regression (RLWR) method~\cite{Chen_2014_2}. Other statistical learning methods can be discussed but we are not going to further investigate them here. 

Up to now, the terminal value $V(T,s)$ has been discussed in a risk-free setting where the reserve price of a guaranteed impression is computed based only on the rebalanced supply of and demand for impressions in the period $[T, \widetilde{T}]$. However, once guaranteed impressions are sold, the publisher take the risk of payment price movement of impressions and the guaranteed selling will affect other advertisers in RTB. The risk externalities can be measured by the standard deviation of the expected payment with regard to the competition level $\xi$, and the terminal value $V(T,s)$ can be then expressed as 
\[
V(T,s) 
= \Bigg\{
\hspace{-5pt} 
\begin{array}{ll}
s \big(\phi(\xi) + \lambda \psi(\xi) \big), & \textrm{if } \pi(\xi) \geq \phi(\xi) + \lambda \psi(\xi),\\
s \pi(\xi), & \textrm{if } \pi(\xi) < \phi(\xi) + \lambda \psi(\xi),\\
\end{array} 
\]
where $\psi(\cdot)$ and $\pi(\cdot)$ are functions which compute the standard deviation of payment and the expected winning bid for the given $\xi$, respectively, and $\lambda$ is the level of risk aversion of the publisher. 

\subsection{Revenue Analysis}

In the following discussion, $\mathbf{R}_{\mathbf{RTB}}$ denotes the expected total revenue of selling all $S$ impressions in RTB, then
\[
\mathbf{R}_{\mathbf{RTB}} = S \phi(Q/S),
\] 
and $\mathbf{R}_{\mathbf{PG+RTB}}$ denotes the expected total revenue of selling some impressions in advance through guaranteed contracts and selling the remaining impressions in RTB, then 
\begin{align*}
\mathbf{R}_{\mathbf{PG+RTB}} = & \ \sum_{t=0}^{T} R(t) x(t) + \Big(S - \sum_{t=0}^{T} x(t) \Big) \phi(\xi^*),
\end{align*}
where $\xi^* = \frac{Q - \sum_{t=0}^{T} x(t)}{S - \sum_{t=0}^{T} x(t)}$.

Let $\mathcal{M}$ be a set such that $x(t) = 1$ for all $t \in \mathcal{M}$. If $\mathcal{M} = \emptyset$, then $\mathbf{R}_{\mathbf{PG+RTB}} = \mathbf{R}_{\mathbf{RTB}}$; if $\mathcal{M} \neq \emptyset$, then
\begin{align*}
  & \ \mathbf{R}_{\mathbf{PG+RTB}}  \\
= & \ 
\sum_{t \in \mathcal{M}} 
G(t) (1- \gamma \omega) x(t) 
+ \Big(S - \sum_{t=0}^{T} x(t) \Big) \phi(\xi^*) \\
\geq & \ 
\sum_{t \in \mathcal{M}} 
r\big(t, S - \hspace{-3pt}\sum_{k \in [0, t]} \hspace{-3pt} x(k)\big) (1- \gamma \omega) 
+ \Big(S - \sum_{t=0}^{T} x(t) \Big) \phi(\xi^*) \\
= & \ 
\sum_{t \in \mathcal{M}} 
\Big(V\big(t, S - \hspace{-3pt}\sum_{k \in [0, t)} \hspace{-3pt} x(k)\big) - V\big(t, S - \hspace{-3pt}\sum_{k \in [0, t]} \hspace{-3pt} x(k)\big) \Big) \\
  & \ + \Big(S - \sum_{t=0}^{T} x(t) \Big) \phi(\xi^*) \\
= & \ 
S \phi(Q/S) + S \lambda \psi(Q/S) - \Big(S-\sum_{t=0}^{T} x(t) \Big) \lambda \psi(\xi^*).  
\end{align*}
If $\lambda = 0$, then $\mathbf{R}_{\mathbf{PG+RTB}} \geq \mathbf{R}_{\mathbf{RTB}}$. If $\lambda > 0$ and $z(\varsigma)^\prime \geq 0$, then $\mathbf{R}_{\mathbf{PG+RTB}} \geq \mathbf{R}_{\mathbf{RTB}}$, where $z(\varsigma) = \varsigma \psi((Q-S+\varsigma)/\varsigma)$. 

Although the model is not optimal, we have shown in which conditions that $\mathbf{R}_{\textbf{PG+RTB}}$ is equal to or higher than $\mathbf{R}_{\mathbf{RTB}}$. Therefore, the publisher' expected revenue can be increased. A proper adjusting the publisher's risk preference will also encourage the model to increase the expected revenue to advertiser's expected value -- the upper bound of any sales model. However, the higher reserve prices may also reject many buy requests, and encourage advertisers to join RTB in the delivery period. In essence, the model won't affect the demand of guaranteed buy requests directly, it is revenue management is based on inventory dynamic allocation.     

\section{Experiments}
\label{sec:experiments}

In this section, we describe our datasets, investigate RTB campaigns, discuss the estimation of model parameters, and evaluate the model's revenue performance.

\subsection{Data and Experimental Design}
\label{sec:data}

\begin{table}[tp]
\centering
\caption{RTB datasets.}
\label{tab:datasets}
\vspace{-5pt}
\begin{tabular}{rlllll}
Dataset & SSP-01 & SSP-02 & DSP\\
\hline
Market  & UK & UK & China \\
From    & 08 Jan 2013 & 01 Jan 2014 & 19 Oct 2013\\
To      & 14 Feb 2013 & 07 Jan 2014 & 27 Oct 2013\\
No. of ad slots    & 31 & 14     & 53571  \\
No. of user tags   & NA & 16600  & 69     \\
No. of publishers  & NA & 5932   & NA     \\
No. of advertisers   & 374 & NA  & 4 \\ 
No. of impressions   & 6646643   & 7752546  & 3158171\\ 
No. of bids          & 33043127  & 7752546 & 11457419 \\
Bid quote               & GBP/CPM   & GBP/CPM  & CNY/CPM\\
Bids of each auction & $\surd$ & NA & NA\\
Reserve price   & NA      & $\surd$  & NA\\
Winning bid     & $\surd$ & $\surd$  & $\surd$\\
Winning payment & $\surd$ & $\surd$  & $\surd$\\
\end{tabular}
\end{table}

\begin{figure}[t]
\centering
\includegraphics[width=1\linewidth]{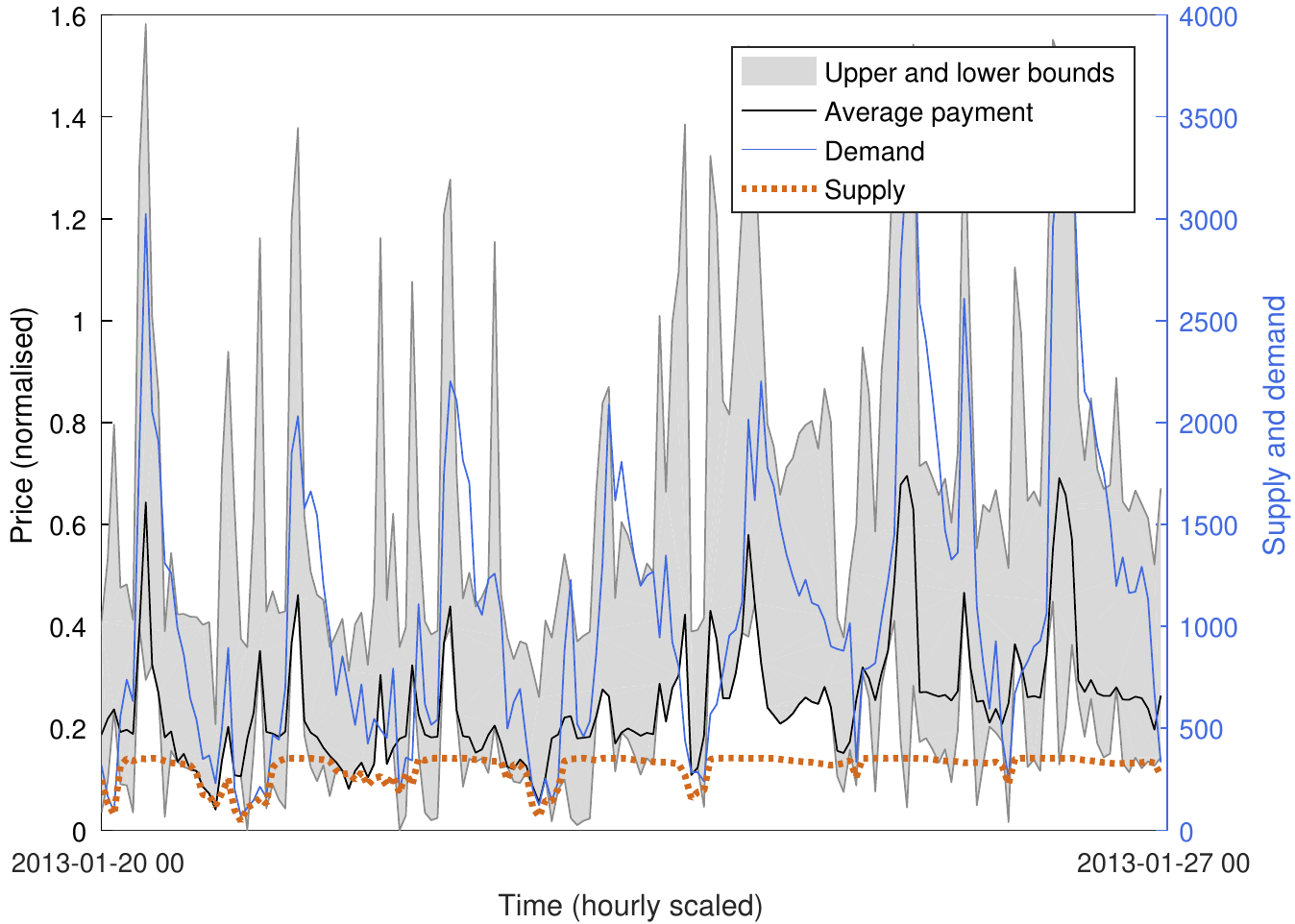}
\caption{An empirical example of RTB campaigns (at hourly time scale) of an ad slot in the SSP-01 dataset.}
\label{fig:demand_payment_movement}
\end{figure}

\begin{table}[btp]
\centering
\caption{Evaluation statistics of surface regression models.}
\label{tab:surface_regression_models_comparison}
\vspace{-5pt}
\begin{tabular}{rllll}
		  & \multicolumn{2}{c}{Demand} & \multicolumn{2}{c}{Supply}\\
		  \cline{2-5}  
Model     &  $L^2$ norm & $L^2$ norm & $L^2$ norm & $L^2$ norm  \\
		  & average & std. & average & std. \\
\hline
PNR(5,5)   &     0.2887  &  0.2544  &  0.2235  &  0.2375\\
PNR(4,5)   &     0.1123  &  0.0824  &  0.0938  &  0.0662\\
PNR(3,5)   &     0.0875  &  0.0507  &  0.0786  &  0.0503\\
PNR(2,5)   &     0.0623  &  0.0435  &  0.0482  &  0.0285\\
PNR(1,5)   &    \cellcolor{Gray}0.0449  &  \cellcolor{Gray}0.0309  &  0.0441  &  0.0276\\
PNR(5,4)   &     0.2979  &  0.2534  &  0.2207  &  0.2379\\
PNR(4,4)   &     0.0874  &  0.0661  &  0.0605  &  0.0434\\
PNR(3,4)   &     0.0856  &  0.0494  &  0.0737  &  0.0493\\
PNR(2,4)   &     0.0597  &  0.0415  &  \cellcolor{Gray}0.0406  & \cellcolor{Gray}0.0234\\
PNR(1,4)   &     0.0496  &  0.0338  &  0.0431  &  0.0271\\
PNR(5,3)   &     0.3024  &  0.2538  &  0.2248  &  0.2374\\
PNR(4,3)   &     0.0877  &  0.0662  &  0.0607  &  0.0433\\
PNR(3,3)   &     0.0736  &  0.0455  &  0.0735  &  0.0517\\
PNR(2,3)   &     0.0579  &  0.0394  &  0.0447  &  0.0256\\
PNR(1,3)   &     0.0476  &  0.0332  &  0.0453  &  0.0280\\
PNR(5,2)   &     0.3061  &  0.2546  &  0.2346  &  0.2368\\
PNR(4,2)   &     0.0896  &  0.0674  &  0.0680  &  0.0456\\
PNR(3,2)   &     0.0789  &  0.0490  &  0.0767  &  0.0534\\
PNR(2,2)   &     0.0622  &  0.0417  &  0.0462  &  0.0267\\
PNR(1,2)   &     0.0529  &  0.0369  &  0.0471  &  0.0273\\
PNR(5,1)   &     0.2807  &  0.2562  &  0.2401  &  0.2340\\
PNR(4,1)   &     0.0880  &  0.0691  &  0.0651  &  0.0438\\
PNR(3,1)   &     0.0804  &  0.0483  &  0.0761  &  0.0538\\
PNR(2,1)   &     0.0672  &  0.0430  &  0.0478  &  0.0310\\
PNR(1,1)   &     0.0566  &  0.0377  &  0.0480  &  0.0307\\
LQR        &     0.0592  &  0.0354  &  0.0546  &  0.0363\\
\end{tabular}
\end{table}

\begin{figure}[t]
\centering
\includegraphics[width=0.8\linewidth]{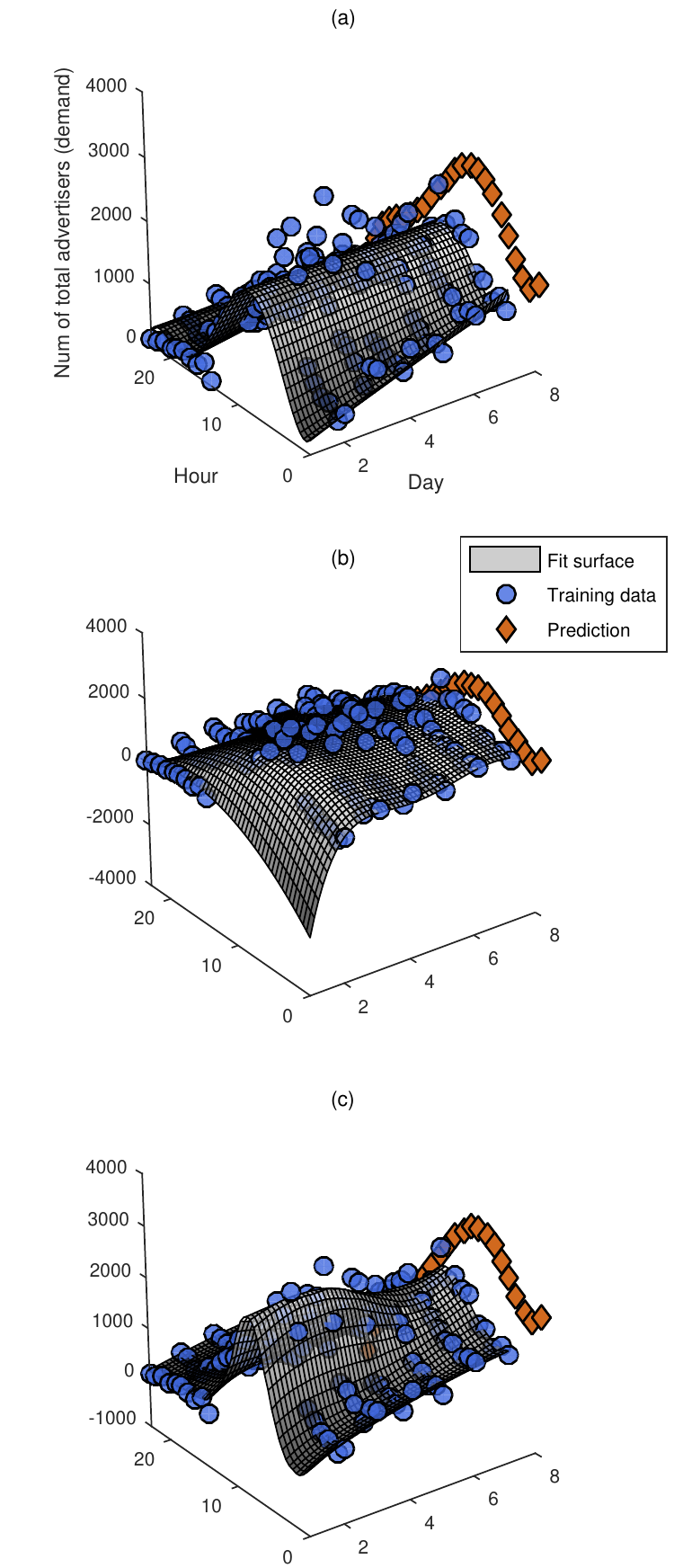}
\caption{Empirical examples of using surface regression models to predict demand in the delivery day for an ad slot in the SSP-01 dataset: (a) PNR(5,5); (b) PNR(2,3); (c) LQR. }
\label{fig:demand_prediction}
\end{figure}

\begin{figure}[t]
\centering
\includegraphics[width=1\linewidth]{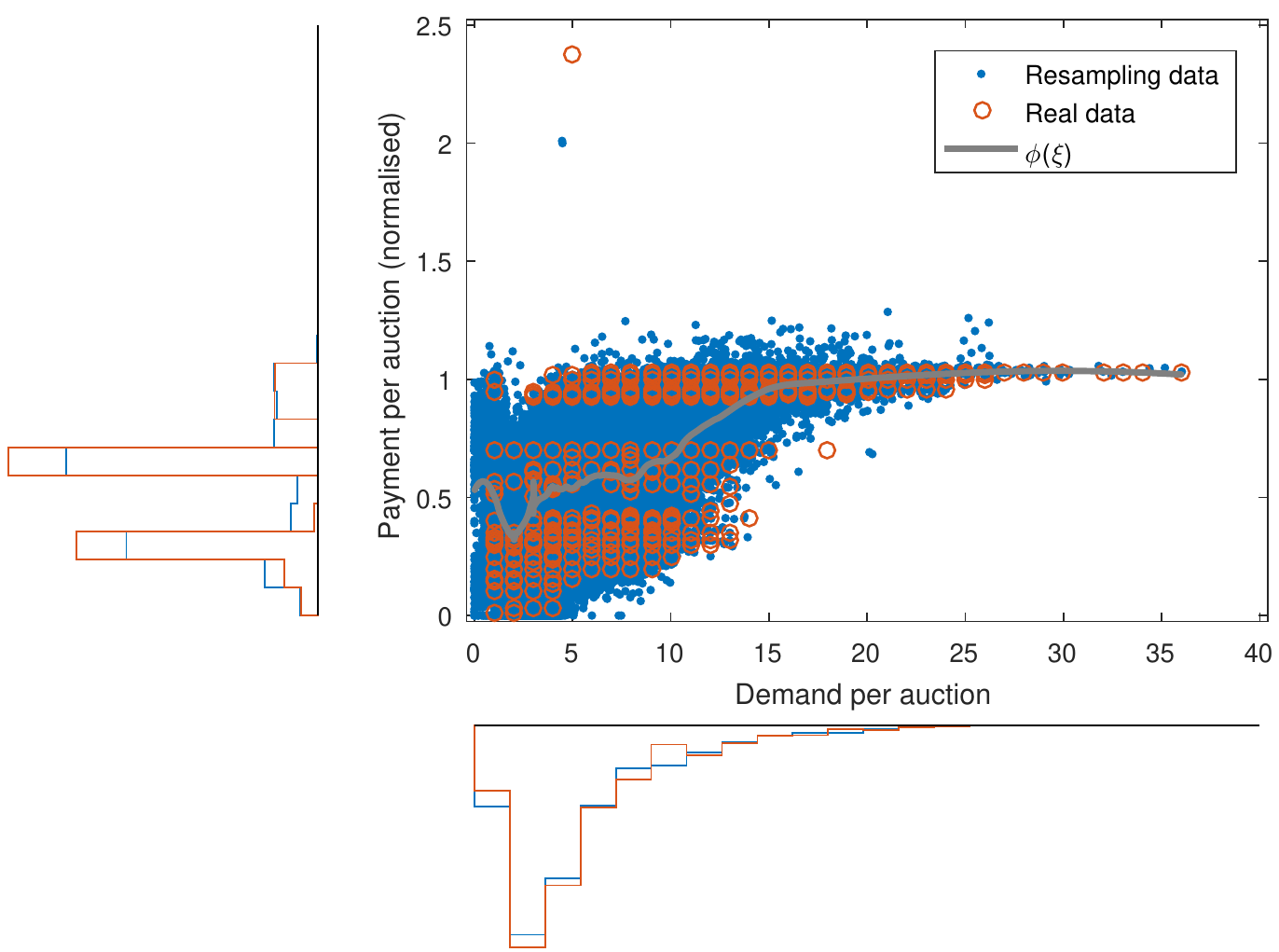}
\caption{An empirical example of estimating $\phi(\xi)$ for an ad slot in the SSP-01 dataset where the resampling rate is 3/2.}
\label{fig:demand_vs_payment}
\end{figure}

Table~\ref{tab:datasets} briefly summarises the used datasets: two datasets from a SSP in the UK over the periods from 08 Jan 2013 to 14 Feb 2013, and from 01 Jan 2014 to 07 Jan 2014; and a dataset from a DSP in China over the period from 19 Oct 2013 to 27 Oct 2013. These datasets contain different information of RTB campaigns; therefore, they are used differently in experiments. The SSP-01 dataset is used throughout the whole experiments and other two datasets are used for further exploring the payment's statistics in RTB. In all datasets, bids are expressed as cost-per-mille (CPM) -- the measurement corresponds to the value of 1000 impressions. 

The SSP-01 dataset contains 31 ad slots and their transactions cover from 8 days to 20 days. Three ad slots (i.e., Slot-25, Slot-26 and Slot-28) have only 8 days (172 consecutive hours) bidding data, and each contains about 58000 auctions. There are 10 ad slots have 20 days (447 consecutive hours) bidding data, and each contains about 130000 to 140000 auctions. In experiments, we randomly select a delivery day for an ad slot so the campaigns reported from that day will be used as the test set and guaranteed contracts can be requested and sold 7 days in advance. Therefore, for the slots with only 8 days data, their delivery days are set to be the 8th day. The prediction of the supply of and demand for impressions in the delivery period will be discussed next. However, it should be noted that forecasting is not our primary intention of this paper. Therefore, we choose a learning period that is close to the delivery day so that the estimated parameters will be more accurate for the evaluation purpose.

\subsection{Parameters Estimation}

Several parameters of the proposed model, such as $Q$, $S$, $\xi$ and $\phi(\cdot)$, whose values can be learned from data. Fig.~\ref{fig:demand_vs_payment} presents an empirical example about the evolution of payment price, supply and demand over time, which is based on a hourly scale. The time series show obvious periodical patterns, and we see a high relevance between payment and the levels of supply and demand. The peak period is between 8:00 and 14:00 every day. This finding is not surprising and is not difficult to explain because these hours lie in the normal working hours and there are a lot of usage of computers and Internet, and therefore, generating a large volume of site visits. However, this periodical pattern will be a little difficult to predict if we only consider a single time variable because the time series change based on a 24-hour cycle. 

In experiments, we divide the time effect into two components: daily effect and hourly effect. Hence, two time variables are considered for prediction. Fig.~\ref{fig:demand_prediction} illustrates the surface regressions that we perform on the training data of an ad slot to predict the next day's demand. We mainly test two types of regression models: (1) polynomial regression (PNR)~\cite{Rawlings_1989}; local quadratic regression (LQR)~\cite{Cleveland_1979}. Table~\ref{tab:surface_regression_models_comparison} presents the evaluation results of the surface regression models on the training data, where PNR(1,5) and PNR(2,4) perform best for demand and supply prediction, respectively. 

To learn $\phi(\cdot)$, we use the RLWR model~\cite{Chen_2014_2}. Fig.~\ref{fig:demand_vs_payment} illustrates an empirical example of estimating the expected per-impression payment given the competition level $\xi$. In the regression, we resample the data at the rate 3/2. It is worth noting that: (1) $\phi(\cdot)$ and $\xi$ are not linearly correlated; (2) a higher competition level will give a higher expected payment price, and this price will converge to a certain level. However, we can not obtain a monotone curve for $\phi(\cdot)$ -- sometimes, a lower competition level will give a slightly higher expected payment. This is because RTB campaigns have been contributed by different advertisers and they have different values on the same impression. The general payment-demand pattern is consistent with our intuitive understanding and the auction theory. In experiments, we learn $\psi(\cdot)$ and $\pi(\cdot)$ in the same way.

\subsection{Results}

\begin{figure}[htp]
\centering
\includegraphics[width=0.93\linewidth]{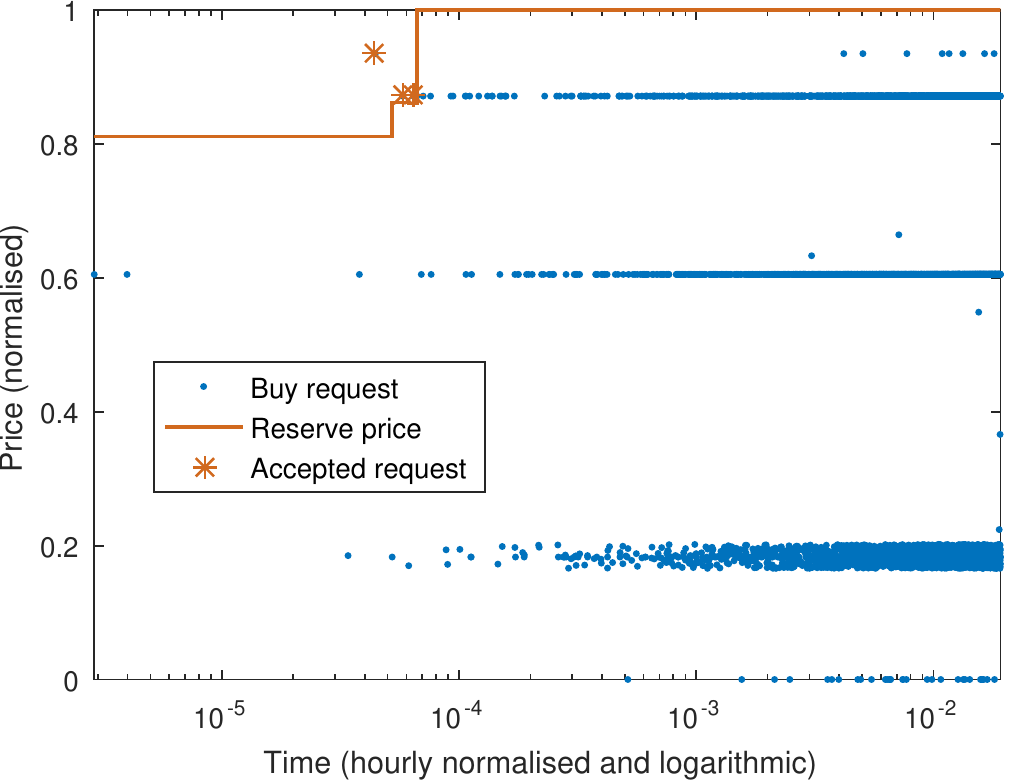}
\caption{An empirical example from an ad slot in the SSP-01 dataset where only a few guaranteed buy requests has been accepted.}
\label{fig:reserve_price_1}\vspace{10pt}
\includegraphics[width=0.93\linewidth]{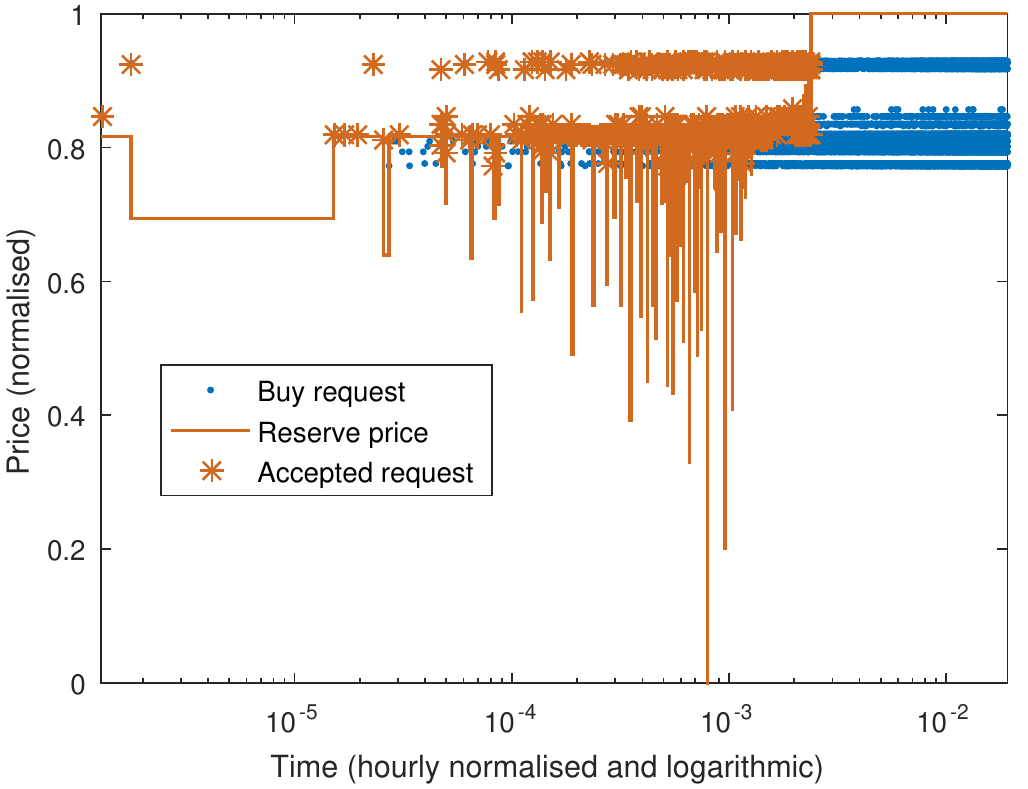}
\caption{An empirical example from an ad slot in the SSP-01 dataset where several guaranteed buy requests have been accepted.}
\label{fig:reserve_price_2}\vspace{10pt}
\includegraphics[width=0.93\linewidth]{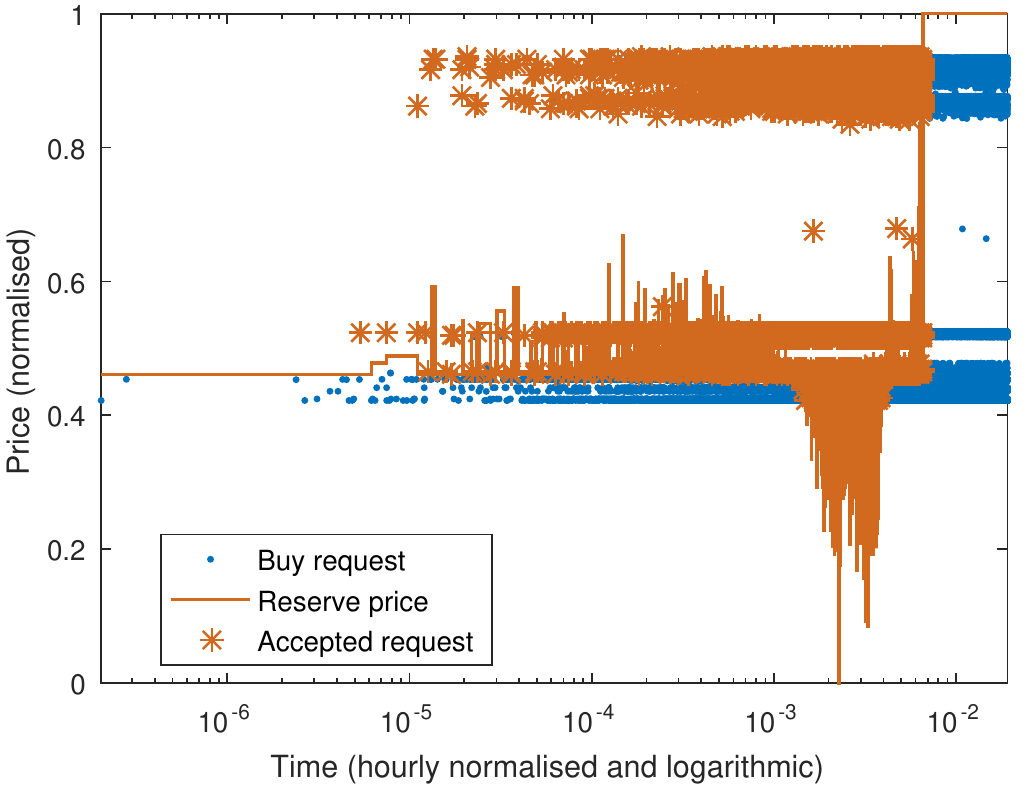}
\caption{An empirical example from an ad slot in the SSP-01 dataset where many guaranteed buy requests have been accepted.}
\label{fig:reserve_price_3}
\end{figure}

In experiments, the guaranteed buy requests are modelled by a homogeneous Poisson process at the intensity rate $Q T$. This rate can be time-dependent. However, as mentioned in Section~\ref{sec:the_model}, the demand arrival won't affect the truth that the proposed model can increase the publisher's revenue. The worst case is that no impression has been sold in advance, and the publisher will auction off all impressions in RTB once they are created. Table~\ref{tab:payment_stats_rtb} shows the payment statistics of winning advertisers in RTB on three datasets. As the advertiser's bid represents her value in the SP auction, the ratio of payment to value shows how well RTB differentiates advertisers and if there is any room to increase revenue through guaranteed contracts. Figs.~\ref{fig:reserve_price_1}-\ref{fig:reserve_price_3} present empirical examples on how the guaranteed buy requests are accepted based on the dynamic reserve prices. In Fig.~\ref{fig:reserve_price_1}, the total revenue is just slightly increased as only a few impressions are sold through guaranteed contracts. Figs.~\ref{fig:reserve_price_2}-\ref{fig:reserve_price_3} show the situations when more buy requests are accepted and how they affect the evolution of reserve prices. The distance between the price of buy request and the reserve price is the revenue increase for an impression. Recall that we have discussed the non-linearity and non-monotonicity relationship between $\phi(\cdot)$ and $\xi$, here we can see the reserve prices are not always increasing or decreasing over time. This is different to active PG~\cite{Chen_2014_2}  or airline tickets booking~\cite{Talluri_2004} where prices over time have a monotone pattern.    

The overall results of revenue performance are summarised in Table~\ref{tab:reves}. Apart from estimating the model parameters using the training data, we also use data in the delivery period, for example, using actual bids to estimate bid distribution, using the actual total supply and demand to estimate the per-impression demand at the given guaranteed price. If we use data in the delivery period, the predicted RTB revenue $\mathbf{R}_{\textbf{RTB}}^{\textbf{Predict}}$ would be very close to, however, in our case is slightly smaller than that from real data $\mathbf{R}_{\textbf{RTB}}$. This shows the variation of the model's approximation and there are 4 slots without revenue growth. In other cases, the revenues are all increasing, validating our revenue analysis. The prediction of future supply and demand is important for the model's performance, which can be further investigated in future research.

\begin{table}[t]
\centering
\caption{Payment statistics of winning advertisers in RTB.}
\label{tab:payment_stats_rtb}
\begin{tabular}{rlll}
		  & Number of        & Ratio of payment  & Ratio of reserve\\
Dataset   & advertisers      & to winning bid    & price to payment\\ 
\hline
SSP-01    & 374     & 51.44\%  & NA \\
SSP-02    & NA      & 77.09\%  & 0.01\%\\
DSP       & 4       & 30.24\%  & NA \\
\end{tabular}
\vspace{7pt}
\caption{Expected revenues.}
\label{tab:reves}
\begin{tabular}{rlll}
 			& Using data in the    	   & Learning data in \\
  			& delivery period          & the training period\\
\hline
$\mathbf{R}_{\textbf{PG+RTB}}^{\textbf{Predict}} \geq \mathbf{R}_{\textbf{RTB}}^{\textbf{Predict}}$  & 100\% & 100\% \\
$\mathbf{R}_{\textbf{PG+RTB}}^{\textbf{Predict}} \geq \mathbf{R}_{\textbf{RTB}}^{\textbf{Real}}$  & 80.77\% & 100\% \\
$(\mathbf{R}_{\textbf{RTB}}^{\textbf{Predict}} - \mathbf{R}_{\textbf{RTB}}^{\textbf{Real}})/\mathbf{R}_{\textbf{RTB}}^{\textbf{Real}}$
& -0.07 & 26.17\\
\end{tabular}
\vspace{-5pt}
\end{table}

\section{Conclusion}
\label{sec:conclusion}

This paper discusses a computational framework of selling a publisher's impressions through guaranteed contracts for display advertising. An advertiser can submit a guaranteed buy request and the publisher makes decisions of whether to sell an impression in advance based on a hidden reserve price. The model assumes static supply and demand in the future delivery period, and the decision making at each request is based on the updated rebalanced supply and demand. We show that the model can increase the publisher's revenue compared to only selling impressions in RTB and validate the the model with SSP datasets as well as discuss in details how to estimate the model parameters.

\bibliographystyle{IEEEtran}
\bibliography{mybib}

\end{document}